\documentclass[reprint, showpacs, preprintnumbers, amsmath,amssymb, aps]{revtex4-1}
\usepackage[colorlinks=true,linkcolor=black, citecolor=black,
urlcolor=black]{hyperref}
\usepackage{txfonts}
\usepackage{adjustbox}

\usepackage{multirow,graphics}

\usepackage{amstext}
\usepackage{amssymb}
\usepackage{amsmath}
\usepackage{graphicx}

\usepackage[usenames,dvipsnames]{color}

\begin{document}

\newcommand{\IM}{{\rm Im}\,}
\newcommand{\card}{\#}
\newcommand{\la}[1]{\label{#1}}
\newcommand{\eq}[1]{(\ref{#1})}
\newcommand{\figref}[1]{Fig \ref{#1}}
\newcommand{\abs}[1]{\left|#1\right|}
\newcommand{\comD}[1]{{\color{red}#1\color{black}}}
\renewcommand{\d}{\partial}

\newcommand{\calN}{\mathcal{N}}\newcommand{\calM}{\mathcal{M}}
\newcommand{\calS}{\mathcal{S}}\newcommand{\calF}{\mathcal{F}}
\newcommand{\calZ}{\mathcal{Z}}\newcommand{\calC}{\mathcal{C}}
\newcommand{\calA}{\mathcal{A}}\newcommand{\calO}{\mathcal{O}}\newcommand{\fQ}{\textsf{Q}}

%%%%%%%%%%%%%%%%%%%%%%%%%%%%%%%%%%%%%%%%%%%%%%%%%%%
%Brackets
%%%%%%%%%%%%%%%%%%%%%%%%%%%%%%%%%%%%%%%%%%%%%%%%%%%
\newcommand{\bra}[1]{\ensuremath{\left< #1\,\right|}}
\newcommand{\ket}[1]{\ensuremath{\left|\, #1\right>}}
\newcommand{\vac}[1]{\ensuremath{\left< \, #1\, \right>}}
%%%%%%%%%%%%%%%%%%%%%%%%%%%%%%%%%%%%%%%%%%%%%%%%%%%

\preprint{DESY 14 - 091, 
 \ HU-Mathematik-14-13, 
 \ HU-EP-14/23}
%\pacs{Whatever}

%%%%%%%%%%%%%%%%%%%%%%%%%%%%
\title{
The Exact Effective Couplings of 4D $\mathcal{N}=2$ gauge theories }
%%%%%%%%%%%%%%%%%%%%%%%%%%%%

\author{ Vladimir Mitev$^{a}$}
\email{mitev@math.hu-berlin.de}
\author{Elli Pomoni$^{b,c}$}
\email{elli.pomoni@desy.de}

\affiliation{
\(^{a}\)Institut f\"ur Mathematik und Institut f\"ur Physik, Humboldt-Universit\"at zu Berlin, IRIS Haus, Zum Gro{\ss}en Windkanal 6,  12489 Berlin, Germany
\\
\(^{b}\)DESY Theory Group, Notkestra{\ss}e 85, 22607 Hamburg, Germany
\\
\(^{c}\)Physics Division, National Technical University of Athens,
15780 Zografou Campus, Athens, Greece             }

%%%%%%%%%%%%%%%%%%%%%%%%%%%%
\begin{abstract}
The anomalous dimensions of operators in the purely gluonic SU(2,1$|$2) sector of any planar conformal $\calN=2$ theory can be read off from the $\calN=4$ SYM results by replacing the $\calN=4$ coupling constant by an interpolating function of the $\calN=2$ coupling constants \cite{Pomoni:2013poa}, to which we refer to as the effective coupling. 
 For a large class of $\calN=2$ theories we compute the weak coupling expansion of  these  functions as well as the leading strong coupling term by employing supersymmetric localization.
Via Feynman diagrams, we interpret our results  as the relative (between  $\calN=2$ and  $\calN=4$) finite renormalization of the coupling constant. Using the AdS/CFT dictionary, we identify the effective couplings with the effective string tensions of the corresponding gravity dual theories.
Thus, any observable  in the SU(2,1$|$2) sector can be obtained from its $\calN=4$ counterpart  by replacing the $\calN=4$ coupling constant by the universal, for a given theory, effective coupling.

\end{abstract}

 \maketitle
%%%%%%%%%%%%%%%%%%%%%%%%%%%%
\section{Introduction and Summary}
%%%%%%%%%%%%%%%%%%%%%%%%%%%%

The recent studies of $\calN=4$ SYM have lead to impressive exact results and novel insights for 4D gauge theories. In this letter we consider the  simplest next step in 4D: $\calN=2$ gauge theories. So far, exact results in gauge theories have come from using either integrability (see \cite{Beisert:2010jr} for a review), localization \cite{Pestun:2007rz}  or a dual string theory description (AdS/CFT \cite{Aharony:1999ti}).

The general problem of obtaining the gravity dual of $\calN=2$ superconformal gauge theories has been studied in \cite{Grana:2001xn,Lin:2004nb, Gaiotto:2009gz,Gadde:2009dj,ReidEdwards:2010qs,Colgain:2011hb,Aharony:2012tz,Stefanski:2013osa} with partial success.  However, theories that are obtained as orbifolds of  $\calN=4$ SYM have well known gravity duals \cite{Kachru:1998ys,Lawrence:1998ja} and in particular, the $\hat{A}_{r-1}$  quivers are dual to $AdS_5\times S^5/\mathbb{Z}_r$, where the $\mathbb{Z}_r$ does not affect the $AdS_5\times S^1$ factor. 
The dual geometry of any $\mathcal{N} = 2$ superconformal theory has an $AdS_5\times S^1$ factor, since the protected members of the  $\mathcal{N}=2$ chiral ring  precisely match the Kaluza-Klein reduction of the 6D Tensor Multiplet on this $AdS_5\times S^1$ factor \cite{Gukov:1998kk,Gadde:2009dj}. 
Wilson loops provide a way to probe the dual geometry and in particular to measure the size of the $AdS_5\times S^1$ factor because, on the string theory side, they are described by a minimal surface which classically ends on the contour of the Wilson loop.
Calculating the expectation value  of the circular Wilson loop on both sides of the correspondence has been one of the first successful tests of the $\mathcal{N}= 4$ AdS/CFT paradigm   \cite{Erickson:2000af,Drukker:2000rr} and with this letter we  begin a similar program for $\mathcal{N}=2$ theories.

In 4D, $\calN=4$ SYM is the unique, up to a choice of the gauge group, maximally supersymmetric gauge theory and it has exactly one marginal coupling constant. The space of conformal $\calN=2$ gauge theories is classified by ADE  \cite{Howe:1983wj,Katz:1997eq,Lawrence:1998ja,Nekrasov:2012xe} finite or affine Dynkin diagrams.  By sending some coupling constant to zero, one can obtain the
superconformal theories that correspond to finite Dynkin diagrams from the affine ones. For simplicity, in the present article we will only consider the elliptic quivers based on the affine $\hat{A}_{r-1}$ Dynkin diagrams that can be obtained from $\mathbb{Z}_r$ orbifolds of  $\calN=4$ SYM. The simplest example in this class is the $\mathbb{Z}_2$ elliptic  quiver. This is the SU$(N_c)\times$SU$(N_c)$ theory with two marginal couplings $g,\check g$ which,  in the limit $\check g\rightarrow 0$, leads to superconformal QCD (SCQCD) with color group SU$(N_c)$ and $N_f=2N_c$ flavor hypermultiplets that has been studied extensively in  \cite{Gadde:2009dj,Gadde:2010zi,Gadde:2010ku,Pomoni:2011jj,Liendo:2011xb,Gadde:2012rv}.

In \cite{Pomoni:2013poa} we show that the purely gluonic SU(2,1$|$2) sector of composite operators in every $\mathcal{N}= 2$ theory,  made out of fields only in the vector multiplet $\phi$, $\lambda_{+}^{\mathcal{I}}$, $\mathcal{F}_{++}$, $\mathcal{D}_{+\dot{\alpha}}$, is closed to all loops in planar perturbation theory. This sector includes operators that correspond to
string states  classically living only on the $AdS_5\times S^1$ factor of the dual geometry. 
We also present a diagrammatic argument that anomalous dimensions  in the SU(2,1$|$2) sector can be read off from the $\calN=4$ ones up to a redefinition, due to finite renormalization, of the coupling constant $g^2\rightarrow f(g^2)$, i.e.
\begin{equation}
\label{anomalousDimReplacement}
\gamma^{\calN=2}(g^2)=\gamma^{\calN=4}(f(g^2))\, ,\quad \text{ where } g^2 = \frac{g^2_{YM}N_c}{(4\pi)^2} \, .
\end{equation}
Thus, we can use the integrability of planar $\calN=4$ and the results available to compute the anomalous dimensions for planar $\calN=2$ theories of operators in this sector, as long as we can compute the effective coupling $f(g^2)$. 

In this letter we compute these functions for the $\hat{A}_{r-1}$ theories \eqref{f-result} and we interpret them as the relative finite renormalization of the coupling constant
\begin{equation}
\label{eq:fasdifference}
 f(g^2)-g^2=g^2\left[\left(\calZ_g^{\calN=2}\right)^2-\left(\calZ_g^{\calN=4}\right)^2\right].
\end{equation}
The calculation of the effective couplings is done via the evaluation of the expectation value of the circular Wilson loop. Using localization,  Pestun  was able to prove the conjecture of  \cite{Erickson:2000af,Drukker:2000rr} that  the  expectation values of the circular Wilson loops for any $\calN=2$ theory can be obtained using matrix models  \cite{Pestun:2007rz}. Here, we use these matrix models to calculate  the Wilson loop expectation values and we show that
\begin{equation}
\label{eq:wilsonloopcouplingreplacement}
W^{\calN=2}(g^2)=W^{\calN=4}(f(g^2)), \text{ with } W^{\calN=4}(g^2)=\frac{I_1(4\pi g)}{2\pi g}  \, .
\end{equation}
From equations \eqref{anomalousDimReplacement} and \eqref{eq:wilsonloopcouplingreplacement} we learn that
the integrable  $\calN=4$  theory knows all about the combinatorics involved in the Feynman diagram calculations.
To get to the  $\calN=2$ theory result all we need to do is to compute  the relative finite renormalization of the coupling constant that is encoded in the effective coupling $f(g^2)$.
On the dual gravity side, the effective couplings are interpreted as the renormalization of the effective string tension 
\begin{equation}
T_{eff}^2= \frac{R^4}{(2 \pi \alpha')^2} = f(g^2)  \,.
\end{equation}
For the $\mathbb{Z}_2$ quiver,  the first correction of the effective coupling $f(g^2)$ from the weak coupling side was computed in \cite{Andree:2010na,Pomoni:2011jj} using Feynman diagrams
\begin{equation}
f(g^2) =  \left\{\begin{array}{ll}g^2 + 12\left(\check{g}^2-g^2\right)\zeta(3) g^4 + \cdots\ ,&  g,\check{g}\rightarrow0\\   2 \frac{g^2 \check{g}^2}{g^2 + \check{g}^2} + \cdots\ , & g,\check{g}\rightarrow \infty\end{array}\right.
\end{equation}
while the  first term of the strong coupling expansion was written in \cite{Gadde:2012rv} by using  AdS/CFT.
In section \ref{WilsonLoopSection}, we write the first few orders of the weak coupling expansion of $f(g^2)$, discuss their Feynman diagram interpretation and give the leading term in the strong coupling limit.

%%%%%%%%%%%%%%%%%%%%%%%%%%%%
\section{The diagrammatic argument and
\\
The power of gauge invariance}
%%%%%%%%%%%%%%%%%%%%%%%%%%%%

Classical gauge theory has local gauge invariance which is broken by the addition of a gauge fixing term during quantization.
The background field formalism (BFF) provides a way to keep manifest as much as possible of the local gauge invariance. To use it, we separate the gauge field $A_{\mu}$ in a classical and a quantum part: $A_{\mu}=\calA_{\mu} +\fQ_{\mu}$. The bare and the renormalized quantities are related by the renormalization factors 
\begin{align}
 & \calA^{\mu}_{\text{bare}}=\sqrt{\calZ_\calA} \calA^{\mu}_{\text{ren}},& &\fQ_{\text{bare}}^{\mu}=\sqrt{\calZ_\fQ} \fQ_{\text{ren}}^{\mu},&\nonumber\\
 &g_{\text{bare}}=\calZ_g g_{\text{ren}},& & \xi_{\text{bare}}=\calZ_{\xi} \xi_{\text{ren}},&
\end{align}
where $\xi$ is the gauge fixing parameter. For simplicity, we present only the Yang Mills part of the theory, but the procedure carries over to quarks and also to supersymmetric $\calN=1$ and $\calN=2$ theories in the appropriate superspace \cite{Gates:1983nr,Buchbinder:1997ya,Buchbinder:1998np,Jain:2013hua}.

In the background field gauge the renormalization factors are related as
\begin{equation}
\calZ_g\sqrt{\calZ_\calA}=1,\quad \calZ_\fQ=\calZ_{\xi}
\end{equation}
and the final answer for any gauge invariant quantity will only depend on the $\calZ_{\calA}$ factor. 
What is more, in the BFF the renormalization factors for the quantum fields $\calZ_\fQ$  will cancel for each individual diagram.
This can be easily seen by recalling a couple of BFF  corollaries.
In the BFF  Feynman diagrams the classical fields $\calA_{\mu}$ cannot propagate on the internal lines. They only appear as external fields in correlation functions. Moreover,
all off-shell n-point functions $\langle \fQ_{\mu_1}\cdots \fQ_{\mu_{\ell_1}} \calA_{\nu_{1}}\cdots \calA_{\nu_{\ell_2}}\rangle$ renormalize as $\calZ^{\ell_1/2}_\fQ \calZ^{\ell_2/2}_\calA \calZ_g^{n} \langle  \fQ_{\mu_1}\cdots \fQ_{\mu_{\ell_1}} \calA_{\nu_{1}}\cdots \calA_{\nu_{\ell_2}}\rangle$. Finally, each internal propagator $\langle \fQ_{\mu} \fQ_{\nu}\rangle$ carries a factor of  $\calZ_\fQ^{-1}$.
Composite local or non-local operators like Wilson loops should be inserted in their renormalized form  $
\mathcal{O}^{\text{ren}}_i\left( \fQ_{\text{ren}} \, , \, \calA_{ren}\right) =\sum_{j} \calZ_{ij}  \mathcal{O}^{\text{bare}}_j\left(\calZ_\fQ^{1/2} \fQ \, , \, \calZ_\calA^{1/2}  \calA \right)
$ where $\calZ_{ij}$ is the  the mixing matrix.

In \cite{Pomoni:2013poa} we presented a diagrammatic argument that for any planar and superconformal $\calN=2$ theory, the asymptotic SU(2,1$|$2) Hamiltonian is identical to all loops to that of  $\calN=4$ SYM, up to a redefinition of the coupling constant $g^2 \rightarrow f(g^2)$. Thus, this sector is integrable and anomalous dimensions can be read off from the  $\calN=4$ ones, up to this redefinition.

A refined version of the diagrammatic argument in \cite{Pomoni:2013poa} is reviewed below, based  only on
 \begin{itemize}
 \item gauge invariance (background field method),
\item the chirality  of the  SU(2,1$|$2) sector which makes the non-renormalization theorem of
\cite{Fiamberti:2008sh,Sieg:2010tz} applicable.
\end{itemize}

 To explain the argument, we begin by considering  $\mathcal{N}=2$  theories obtained as  orbifolds of $\mathcal{N}=4$ SYM. They are conformal by inheritance arguments  \cite{Bershadsky:1998mb,Bershadsky:1998cb}.
When all the coupling constants are equal to each other (orbifold point),  all anomalous dimensions in the untwisted sector are equal to the $\mathcal{N}=4$ ones. 

In order to compute the renormalization of operators, we write down all the relevant diagrams and compute each one of them in $\calN=4$ (at the orbifold point) as well as in $\calN=2$ and subtract the results from each other. 
All  the individual UV-divergent Feynman diagrams that should be calculated for the renormalization of operators in the SU(2,1$|$2) sector, are identical in  both theories. The only diagrams that are different from their $\calN=4$ counterparts are finite and they  are responsible for the relative finite renormalization between the $\calN=2$ and the $\calN=4$ coupling constants. Some examples of such diagrams are depicted in figures \ref{fig:zeta3}, \ref{fig:zeta5} and \ref{fig:zeta3square}.

This procedure should be thought of as a novel regularization prescription that cancels the divergencies of each individual diagram. The fact that the difference of the two diagrams is always finite stems from the finiteness of the $\mathcal{N}=2$  theories we are considering \cite{Howe:1983wj} and from the fact that the purely gluonic  tree level terms in both the $\calN=2$ and the $\calN=4$  Lagrangians are identical. With this powerful  regularization prescription, we can simplify our computations. All the combinatorics and symmetry factors of the individual diagrams are identical in both theories. So, we let the $\calN=4$ integrable model give them to us, and we just have to compute the difference \eqref{eq:fasdifference}.

There is one possible way this argument could fail.
Going up to higher order in $g$,  new nonlocal  vertices can appear in the effective action of  $\mathcal{N}=2$  theories that are not there  for  $\mathcal{N}=4$ SYM.  
However, none of these new vertices can  contribute to the anomalous dimensions of the SU(2,1$|$2) sector  \cite{Pomoni:2013poa} due to the non-renormalization theorem of
\cite{Fiamberti:2008sh,Sieg:2010tz}. Only the renormalized tree level vertices will contribute. Due to the fact that the $\calZ_\fQ$ cancel, the final result depends only on $\calZ_\calA= \calZ_g^{-2}$. Thus all anomalous dimensions obey $\gamma_i \big( g^2 \big) = \gamma^{\mathcal{N}=4}_i \big( f(g^2) \big)$ with $f(g^2)$ given in \eqref{eq:fasdifference}.

%%%%%%%%%%%%%%%%%%%%%%%%%%%%
\section{Wilson loops}
\label{WilsonLoopSection}
%%%%%%%%%%%%%%%%%%%%%%%%%%%%

 Pestun's matrix models provide an efficient way to compute the expectation value of the circular Wilson loop 
\begin{equation}
\label{eq:wilsonloopexpectationvalue}
W_{k}^{\calN=2}=\vac{\frac{1}{N_c}\text{tr}_{\square}\text{Pexp}\oint_{\calC}ds \left(i A_{\mu}^{(k)}(x)\dot{x}^{\mu}+\phi^{(k)}(x)|\dot{x}|\right)}\ ,
\end{equation}
where $\square$ denotes the fundamental representation and $\calC$ is the circular loop located at the equator of $S^4$.  The adjoint scalar $\phi^{(k)}$  and the gauge field $A_{\mu}^{(k)}$ are in the vector multiplet of the $k$-th gauge group. Inserting in the path integral a composite operator with fields only in the  $k$-th vector multiplet selects the coupling $g_k^2$ whose renormalization we are computing,
\begin{equation}
\label{eq:wilsonloopcouplingreplacement-k}
W_{k}^{\calN=2}(g_1,\ldots, g_r)=W^{\calN=4}(f_k(g_1,\ldots,g_r))\ ,
\end{equation}
where $f_k(g_1,\ldots,g_r)=g_k^2+\cdots$ is the effective coupling constant of the $k$-th gauge group. 

Let us consider a cyclic quiver made out of $r$ gauge groups, corresponding to the untwisted affine Dynkin diagram $\hat{A}_{r-1}$. We follow the method and notations of \cite{Passerini:2011fe,Russo:2012ay,Russo:2013kea,Russo:2013sba}. The partition function of the corresponding matrix model is
\begin{equation}
\begin{split}
Z=&\int\prod_{k=1}^rda^{(k)}\prod_{i<j=1}^{N_c}\left(a_i^{(k)}-a_j^{(k)}\right)^2 e^{-\frac{N_c}{2g_k^2}\sum_{i=1}^{N_c}\left(a_i^{(k)}\right)^2}Z_{\text{1-loop}}\left|Z_{\text{inst}}\right|^2.
\end{split}
\end{equation}
In the planar limit, the instanton contribution can be neglected, while the one loop part is 
\begin{equation}
Z_{\text{1-loop}}=\prod_{k,l=1}^r\prod_{i,j=1}^NH^{\frac{\textbf{a}_{kl}}{2}}\big(a_i^{(k)}-a_j^{(l)}\big),
\end{equation}
where $H(x)=\prod_{n=1}^{\infty}\left(1+\frac{x^2}{n^2}\right)^ne^{-\frac{x^2}{n}}$ and $\textbf{a}_{kl}$ is the Cartan matrix corresponding to $\hat{A}_{r-1}$.
By using the saddle point approximation and replacing  in the planar limit  the eigenvalues $a_i^{(k)}$ by normalized densities $\rho_k(x)$ that are localized in an interval $[-\mu_k,\mu_k]$, we obtain  the following system of coupled integral equations:
\begin{equation}
\label{eq:saddlepoint}
\frac{x}{2g_k^2}=\fint_{-\mu_k}^{\mu_k}\frac{\rho_k(y)}{x-y}-\frac{1}{2}\sum_{l=1}^r\textbf{a}_{kl}\int_{-\mu_l}^{\mu_l}\rho_l(y)K(x-y)dy,
\end{equation}
for $k=1,\ldots, r$. 
For small values of the couplings the widths of the densities tend towards zero and we can expand the kernel 
$K(x)=-2\sum_{n=1}^{\infty}(-1)^n\zeta(2n+1)x^{2n+1}.$
Then we can solve the integral equations recursively and compute the Wilson loop expectation values of equation \eqref{eq:wilsonloopexpectationvalue} via:
\begin{equation}
W_{k}^{\calN=2}=\vac{\frac{1}{N_c}\sum_{i=1}^{N_c}e^{2\pi a_i^{(k)}}}=\int_{-\mu_k}^{\mu_k}\rho_k(x)e^{2\pi x}dx. 
\end{equation}

For the elliptic $\mathbb{Z}_2$ quiver with couplings $g_1=g$, $g_2=\check{g}$, we obtain
\begin{eqnarray}
\label{eq:fresultZ2}
&&f(g,\check{g})=g^2+ 2\left(\check{g}^2-g^2\right)\Big[6\zeta(3) g^4 -20 \zeta(5) g^4\left(\check{g}^2+3 g^2\right) \nonumber\\&&+ g^4 \Big(70\zeta(7) \Big(\check{g}^4+5\check{g}^2g^2+8g^4\Big)-2 \zeta(2)(20\zeta(5)) g^4 \nonumber\\&&-2(6\zeta(3))^2  \Big(\check{g}^4-\check{g}^2 g^2+2 g^4 \Big)  \Big)\Big]+\cdots
\end{eqnarray}
Inserting the above in $W^{\calN=4}(f(g,\check{g}))$ and taking the limit $\check{g}\rightarrow 0$, we recover the $\calN=2$ SCQCD computation of \cite{Passerini:2011fe}.
For the general superconformal cyclic $\hat{A}_{r-1}$ quivers, we  obtain up to order $\calO(g^{10})$:
\begin{eqnarray}
\label{f-result}
f_k&=&g_k^2+6\zeta(3) g_k^4 \left[g_{k-1}^2+g_{k+1}^2-2 g_k^2\right] \nonumber\\&&-20 \zeta(5) g_k^4 \left[g_{k-1}^4+g_{k+1}^4-6 g_k^4+2 g_k^2 \left(g_{k-1}^2+g_{k+1}^2\right)\right] \nonumber\\&&+ g_k^4 \Big[70\zeta(7) \Big(g_{k-1}^6+g_{k+1}^6-16 g_k^6+3 g_k^4 \left(g_{k-1}^2+g_{k+1}^2\right)\\&&+4 g_k^2 \left(g_{k-1}^4+g_{k+1}^4\right)\Big)-2 \zeta(2)(20\zeta(5)) g_k^4 \left(g_{k-1}^2+g_{k+1}^2-2 g_k^2\right)\nonumber\\&&+(6\zeta(3))^2 \Big(8 g_k^6-2 g_{k-1}^6-2 g_{k+1}^6+g_{k-1}^4 g_{k-2}^2+g_{k+2}^2 g_{k+1}^4\nonumber\\&&-6 g_k^4 \left(g_{k-1}^2+g_{k+1}^2\right)+2 g_k^2 \left(g_{k-1}^4+g_{k-1}^2 g_{k+1}^2+g_{k+1}^4\right)\Big)  \Big]+\cdots\nonumber,
\end{eqnarray}
while the leading term at strong coupling is
\begin{equation}
f_k=r\frac{g_1^2\cdots g_r^2}{\sum_{i=1}^r\prod_{j\neq i}g_j^2}+\cdots,
\end{equation}
which agrees with the AdS/CFT prediction \cite{Lawrence:1998ja,Gadde:2009dj, Gadde:2012rv}. 
The $\mathbb{Z}_r$ symmetry implies the following cyclic relation
\begin{equation}
f_k(g_1,\ldots, g_r)= f_{k+l}(g_{1+l},\ldots, g_{r+l}), \quad \forall k,l,
\end{equation}
i.e. all the effective couplings are given by the same function, up to a cyclic shift of the couplings. 

%%%%%%%%%%%%%%%%%%%%%%%%%%%%%%%%%%%%%%%%%%%%%%%%%%
\section{Feynman diagram interpretation}
%%%%%%%%%%%%%%%%%%%%%%%%%%%%%%%%%%%%%%%%%%%%%%%%%%
Calculating $f(g^2)$  using Feynman diagrams  is not as hard as one would imagine because of its interpretation as the relative finite renormalization of the coupling constant \eqref{eq:fasdifference}. 
First of all, in the BFF one does not have to calculate the renormalization of 3- or 4- point vertices as for usual covariant gauges, but to use $\calZ_g= \calZ_\calA^{-1/2}$ and to compute only the renormalization of the propagator $\vac{\calA(p)\calA(-p)}$.
Moreover, to get \eqref{eq:fasdifference}, we do not need to calculate every single diagram that contributes to the renormalization of the propagator, but only the ones that are different between $\calN=2$ and   $\calN=4$ (or the orbifold of  $\calN=4$ at the orbifold point).
As we discussed in \cite{Pomoni:2013poa, Pomoni:2011jj}, for any
$\calN=2$ 
superconformal theory the only possible way to get diagrams different from the
$\calN=4$ 
ones is to make a loop with hypermultiplets and to let a vector field from  a neighboring vector multiplet propagate inside this loop. This narrows down significantly the number of Feynman diagrams that need to be computed.

\begin{figure}[h]
\begin{centering}
\includegraphics[height=1.7cm]{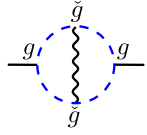}
\end{centering}
\caption{The Feynman diagram responsible for the $\zeta(3)$ contribution to $f(g,\check{g})$. The
solid lines represents the background $\calN=2$ vector superfield, the wiggly line the quantum $\calN=2$ vector superfield while the dashed blue lines  the $\calN=2$ hyperfield.
}
\label{fig:zeta3}
\end{figure}

\begin{figure}[h]
\begin{centering}
\includegraphics[height=1.7cm]{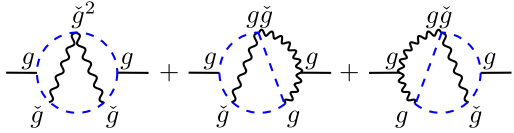}
\end{centering}
\caption{In this figure we present 
some representative Feynman diagrams that are responsible for the $\zeta(5)$ contribution to $f(g,\check{g})$.}
\label{fig:zeta5}
\end{figure}

It so happens that the type of diagrams that are \emph{different} from the
$\calN=4$ ones are always finite and they always include as a basic building block the finite fan integrals of \cite{Broadhurst:1985vq}. For a fan with $n$ faces, we have
\begin{equation}
\label{eq:Broadhurstfan}
\includegraphics[ height=1cm]{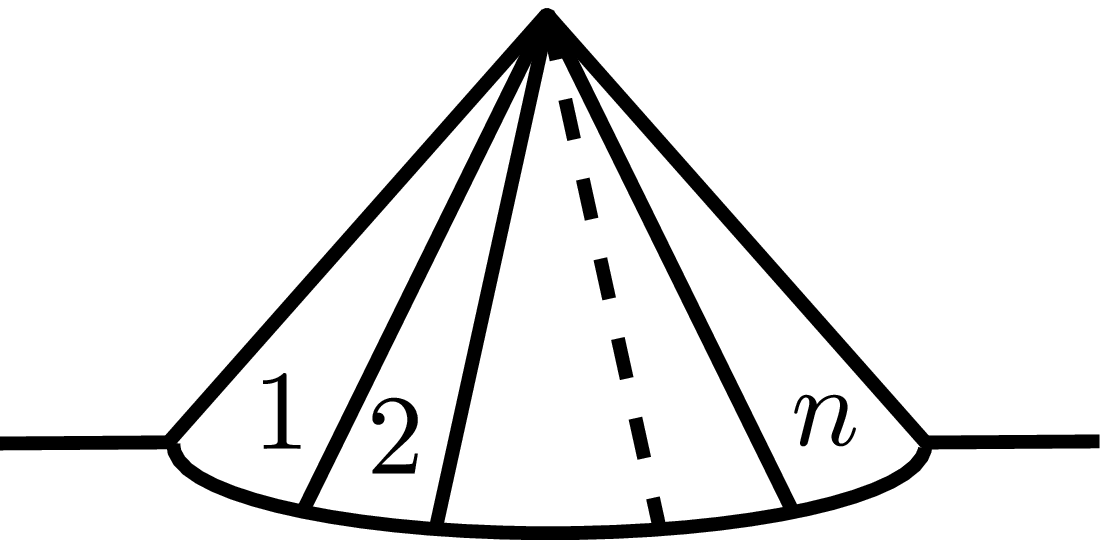}\begin{array}{c} =2\left(\begin{array}{c}2n-1\\n\end{array}\right)\zeta(2n-1)  \frac{1}{p^2}  \, . \\ \\   \end{array}
\end{equation}
The first $\zeta(3)$ contribution in \eqref{eq:fresultZ2} was computed in \cite{Pomoni:2011jj}, it comes from the diagram depicted in figure~\ref{fig:zeta3} and is  equal to $12g^4\check{g}^2\zeta(3)$.
Subtracting from it the $\calN=4$ result of $12g^6\zeta(3)$ gives precisely the $\zeta(3)$ coefficient in \eqref{eq:fresultZ2}. 
The next correction comes with a $\zeta(5)$  and is obtained from diagrams, of different graph topologies, in which two propagators are running in the bubble. For each graph topology there are three diagrams. In figure~\ref{fig:zeta5}, we give an example of the diagrams with the same topology. Their sum is proportional to $\check{g}^4g^4+2\check{g}^2g^6$, which  happens to be the same for all topologies. After subtracting the $\calN=4$ result, which is proportional to $3g^8$, from the $\calN=2$ one  we obtain the overall coefficient of $20\zeta(5)$: $g^4(\check{g}^2-g^2)(\check{g}^2+3g^2)$ which is equal to the one in \eqref{eq:fresultZ2}.
Observe that the overall sign of the $\zeta(2n-1)$ contribution is alternating, because each wiggly line comes with a minus sign.

\begin{figure}[h]
\begin{centering}
\includegraphics[height=2cm]{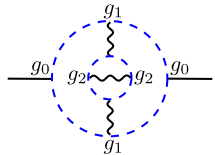}
\end{centering}
\caption{This figure shows a part of the $\zeta(3)^2$ contribution created by nested Feynman diagrams.}
\label{fig:zeta3square}
\end{figure}

All fan diagrams come with maximum transcendentality for the given loop level.
However, as we see in  \eqref{eq:fresultZ2} and \eqref{f-result}, less than maximum transcendentality  contributions can appear. These come from nested diagrams like the one depicted in figure  \ref{fig:zeta3square}.
For the general cyclic quiver with $r>2$, we start getting contributions from the next to nearest neighbor gauge groups for the $\zeta(3)^2$ term of \eqref{f-result}.

Finally, the origin of the $\zeta(2)\zeta(5)$ contribution hasn't been fully elucidated and a careful calculation is in progress.

%%%%%%%%%%%%%%%%%%%%%%%%%%%%
\section{Discussion and Future work}
%%%%%%%%%%%%%%%%%%%%%%%%%%%%

Building on  \cite{Pomoni:2013poa},  we have argued that the anomalous dimensions of operators in the purely gluonic SU(2,1$|$2) sector of conformal $\calN=2$ gauge theories can be obtained by taking the corresponding $\calN=4$ result and replacing the $\calN=4$ coupling constant by the effective coupling $f(g^2)$. 
Localization provides exact results for the expectation values of circular Wilson loops,  from which we determine the weak coupling expansion as well as the leading strong coupling term of the effective couplings.
Finally, we interpreted $f(g^2)$ diagrammatically  as the relative finite renormalization of the coupling constant between the $\calN=2$ and the  $\calN=4$ theories \eqref{eq:fasdifference}. This provides a test of our results using Feynman diagrams.

Based on the existence of  an AdS dual description of these  $\calN=2$ gauge theories and on the interpretation of the effective couplings  as the effective string tensions $T_{eff}^2= f(g^2)$, we conjecture that all possible observables that are restricted to the purely gluonic SU(2,1$|$2) sector  can be computed by replacing  the $\calN=4$ coupling constant in the corresponding results by the {\it universal}  effective coupling $f(g^2)$. Such observables include the cusp anomalous dimension 
\cite{Korchemsky:2010kj}, scattering amplitudes, Wilson loops (see \cite{CaronHuot:2010ek} and references therein) and correlation functions \cite{Eden:2011yp, Eden:2011ku} with the external momenta constrained by $p_{-\dot{\alpha}}=0$.

In a future work, we will present similar results for the asymptotically conformal quiver theories in which conformality is softly broken by adding mass terms for the hypermultiplets. While our methods are applicable and $g^2$ is again corrected only with finite renormalization, understanding the scheme dependence is subtle and requires further investigation.

One way to test our results for the effective couplings is to consider  the anomalous dimension of the twist-two descendant of Konishi. Using the anomalous dimension of Konishi in $\calN=4$ SYM, which thanks to integrability is known up to eight loops  \cite{Leurent:2013mr}, we can predict the anomalous dimension of the  twist-two $D^2Z^2$ descendant to the same loop order for any of the cyclic quivers. 
This prediction can be explicitly checked by computing the wrapping corrections using Feynman diagram calculations to at least four loops following \cite{Fiamberti:2007rj}.
To conserve space, we restrict ourselves to five loops and to the $\hat{A}_{1}$ quiver. The contributions to $\Delta$ that differ from the $\calN=4$ ones are marked in red:
\begin{eqnarray}
&&\Delta(g,\check{g})=4+12 g^2 -48 g^4 +48g^4  \left[7g^2 {\color{red}{-3 \left(g^2-\check{g}^2\right) \zeta(3)}}\right]\nonumber\\&&+96g^4 \Big[ -26g^4+6 \zeta(3)g^4-15\zeta(5) g^4{\color{red}{+\left(g^2-\check{g}^2\right)\Big(12 g^2  \zeta(3)}}\nonumber\\&&{\color{red}{+5 \left(3g^2+\check{g}^2\right)\zeta(5)\Big)}}\Big]+16 g^4 \Big[948g^6+432g^6\zeta(3)\nonumber\\&&-324g^6\zeta(3)^2-540g^6\zeta(5)+1890 g^6\zeta(7)
\\&&
{\color{red}{-3\left(g^2- \check{g}^2\right)\Big[
\left(8 g^4+5 g^2 \check{g}^2+\check{g}^4\right) 35 \zeta(7)
}}\nonumber\\&&{\color{red}{- g^2 \left(4 \check{g}^2+g^2 \left(12- \zeta(2)\right)\right)20 \zeta(5)}}\nonumber\\&&{\color{red}{
- \left(2g^4-g^2\check{g}^2+\check{g}^4\right)\left( 6\zeta(3) \right)^2
+ 42g^4 \left( 6\zeta(3) \right)
\Big] }}\Big]+\cdots.\nonumber
\end{eqnarray}

Our results can also be used for observables outside the SU(2,1$|$2) sector. The all loop dispersion relation and scattering matrix for fields in the bifundamental hypermultiplet  in the $\phi$ vacuum 
were derived in \cite{Gadde:2010ku}
\begin{equation}
E_{\mbox{\footnotesize bif}}(p)=\sqrt{1+4\left(\textbf{g}-\check{\textbf{g}}\right)^2+16\textbf{g}\check{\textbf{g}}\sin^2\left(\frac{p}{2}\right)}\, ,
\end{equation}
up to two unknown functions that we conjecture are given by the effective couplings $\textbf{g}=f(g,\check{g})^{\frac{1}{2}}$ and $\check{\textbf{g}}=\check{f}(g,\check{g})^{\frac{1}{2}}$ that we calculated \eqref{eq:fresultZ2} in this paper. Due to the $\mathbb{Z}_2$ symmetry, we have $\check{f}(g,\check{g})=f(\check{g},g)$. Thus  the dispersion and the scattering matrix are now exactly known.

Our work is the 4D equivalent of the ABJM result of \cite{Gromov:2014eha}, even though our methods are very different. The quantum spectral curve and the slope functions of Basso \cite{Basso:2011rs} can be used to check our logic. Furthermore, the results for the interpolating function $h_{ABJM}(g)$ \cite{Gromov:2014eha} can be combined with our method to derive $h_{ABJ}(g)$ of the ABJ theory \cite{Aharony:2008gk}.

The present letter contains the principles regarding the computation of the effective couplings via localization as well as their Feynman diagram interpretation. In a forthcoming publication, we shall provide additional details, shall give expressions concerning the strong coupling expansion and the implementation of mass terms for the hypermultiplets.

%%%%%%%%%%%%%%%%%%%%%%%%%%%%
\section*{Acknowledgments}
\label{sec:acknowledgments}
We are grateful to Wenbin Yan for giving us his unpublished notes on the calculation of the Wilson loop of the $\mathbb{Z}_2$ quiver using localization.
 We thank   
Isabella Bierenbaum, Sergey Frolov, Sven-Olaf Moch, Leonardo Rastelli, Martin Ro\v cek,  Alessandro Sfondrini, Christoph Sieg and Peter van Nieuwenhuizen for insightful and helpful discussions.


\begin{thebibliography}{48}%
\makeatletter
\providecommand \@ifxundefined [1]{%
 \@ifx{#1\undefined}
}%
\providecommand \@ifnum [1]{%
 \ifnum #1\expandafter \@firstoftwo
 \else \expandafter \@secondoftwo
 \fi
}%
\providecommand \@ifx [1]{%
 \ifx #1\expandafter \@firstoftwo
 \else \expandafter \@secondoftwo
 \fi
}%
\providecommand \natexlab [1]{#1}%
\providecommand \enquote  [1]{``#1''}%
\providecommand \bibnamefont  [1]{#1}%
\providecommand \bibfnamefont [1]{#1}%
\providecommand \citenamefont [1]{#1}%
\providecommand \href@noop [0]{\@secondoftwo}%
\providecommand \href [0]{\begingroup \@sanitize@url \@href}%
\providecommand \@href[1]{\@@startlink{#1}\@@href}%
\providecommand \@@href[1]{\endgroup#1\@@endlink}%
\providecommand \@sanitize@url [0]{\catcode `\\12\catcode `\$12\catcode
  `\&12\catcode `\#12\catcode `\^12\catcode `\_12\catcode `\%12\relax}%
\providecommand \@@startlink[1]{}%
\providecommand \@@endlink[0]{}%
\providecommand \url  [0]{\begingroup\@sanitize@url \@url }%
\providecommand \@url [1]{\endgroup\@href {#1}{\urlprefix }}%
\providecommand \urlprefix  [0]{URL }%
\providecommand \Eprint [0]{\href }%
\providecommand \doibase [0]{http://dx.doi.org/}%
\providecommand \selectlanguage [0]{\@gobble}%
\providecommand \bibinfo  [0]{\@secondoftwo}%
\providecommand \bibfield  [0]{\@secondoftwo}%
\providecommand \translation [1]{[#1]}%
\providecommand \BibitemOpen [0]{}%
\providecommand \bibitemStop [0]{}%
\providecommand \bibitemNoStop [0]{.\EOS\space}%
\providecommand \EOS [0]{\spacefactor3000\relax}%
\providecommand \BibitemShut  [1]{\csname bibitem#1\endcsname}%
\let\auto@bib@innerbib\@empty
%</preamble>
\bibitem [{\citenamefont {Pomoni}(2013)}]{Pomoni:2013poa}%
  \BibitemOpen
  \bibfield  {author} {\bibinfo {author} {\bibfnamefont {E.}~\bibnamefont
  {Pomoni}},\ }\href@noop {} {\  (\bibinfo {year} {2013})},\ \Eprint
  {http://arxiv.org/abs/1310.5709} {arXiv:1310.5709} \BibitemShut
  {NoStop}%
%%CITATION = ARXIV:1310.5709;%%
\bibitem [{\citenamefont {Beisert}\ \emph {et~al.}(2012)\citenamefont
  {Beisert}, \citenamefont {Ahn}, \citenamefont {Alday}, \citenamefont
  {Bajnok}, \citenamefont {Drummond} \emph {et~al.}}]{Beisert:2010jr}%
  \BibitemOpen
  \bibfield  {author} {\bibinfo {author} {\bibfnamefont {N.}~\bibnamefont
  {Beisert}}, \bibinfo {author} {\bibfnamefont {C.}~\bibnamefont {Ahn}},
  \bibinfo {author} {\bibfnamefont {L.~F.}\ \bibnamefont {Alday}}, \bibinfo
  {author} {\bibfnamefont {Z.}~\bibnamefont {Bajnok}}, \bibinfo {author}
  {\bibfnamefont {J.~M.}\ \bibnamefont {Drummond}},  \emph {et~al.},\ }\href
  {\doibase 10.1007/s11005-011-0529-2} {\bibfield  {journal} {\bibinfo
  {journal} {Lett.Math.Phys.}\ }\textbf {\bibinfo {volume} {99}},\ \bibinfo
  {pages} {3} (\bibinfo {year} {2012})},\ \Eprint
  {http://arxiv.org/abs/1012.3982} {arXiv:1012.3982 } \BibitemShut
  {NoStop}%
%%CITATION = ARXIV:1012.3982;%%
\bibitem [{\citenamefont {Pestun}(2012)}]{Pestun:2007rz}%
  \BibitemOpen
  \bibfield  {author} {\bibinfo {author} {\bibfnamefont {V.}~\bibnamefont
  {Pestun}},\ }\href {\doibase 10.1007/s00220-012-1485-0} {\bibfield  {journal}
  {\bibinfo  {journal} {Commun.Math.Phys.}\ }\textbf {\bibinfo {volume}
  {313}},\ \bibinfo {pages} {71} (\bibinfo {year} {2012})},\ \Eprint
  {http://arxiv.org/abs/0712.2824} {arXiv:0712.2824 } \BibitemShut
  {NoStop}%
%%CITATION = ARXIV:0712.2824;%%
\bibitem [{\citenamefont {Aharony}\ \emph {et~al.}(2000)\citenamefont
  {Aharony}, \citenamefont {Gubser}, \citenamefont {Maldacena}, \citenamefont
  {Ooguri},\ and\ \citenamefont {Oz}}]{Aharony:1999ti}%
  \BibitemOpen
  \bibfield  {author} {\bibinfo {author} {\bibfnamefont {O.}~\bibnamefont
  {Aharony}}, \bibinfo {author} {\bibfnamefont {S.~S.}\ \bibnamefont {Gubser}},
  \bibinfo {author} {\bibfnamefont {J.~M.}\ \bibnamefont {Maldacena}}, \bibinfo
  {author} {\bibfnamefont {H.}~\bibnamefont {Ooguri}}, \ and\ \bibinfo {author}
  {\bibfnamefont {Y.}~\bibnamefont {Oz}},\ }\href {\doibase
  10.1016/S0370-1573(99)00083-6} {\bibfield  {journal} {\bibinfo  {journal}
  {Phys. Rept.}\ }\textbf {\bibinfo {volume} {323}},\ \bibinfo {pages} {183}
  (\bibinfo {year} {2000})},\ \Eprint {http://arxiv.org/abs/hep-th/9905111}
  {arXiv:hep-th/9905111} \BibitemShut {NoStop}%
%%CITATION = HEP-TH/9905111;%%
\bibitem [{\citenamefont {Grana}\ and\ \citenamefont
  {Polchinski}(2002)}]{Grana:2001xn}%
  \BibitemOpen
  \bibfield  {author} {\bibinfo {author} {\bibfnamefont {M.}~\bibnamefont
  {Grana}}\ and\ \bibinfo {author} {\bibfnamefont {J.}~\bibnamefont
  {Polchinski}},\ }\href {\doibase 10.1103/PhysRevD.65.126005} {\bibfield
  {journal} {\bibinfo  {journal} {Phys.Rev.}\ }\textbf {\bibinfo {volume}
  {D65}},\ \bibinfo {pages} {126005} (\bibinfo {year} {2002})},\ \Eprint
  {http://arxiv.org/abs/hep-th/0106014} {arXiv:hep-th/0106014}
  \BibitemShut {NoStop}%
%%CITATION = HEP-TH/0106014;%%
\bibitem [{\citenamefont {Lin}\ \emph {et~al.}(2004)\citenamefont {Lin},
  \citenamefont {Lunin},\ and\ \citenamefont {Maldacena}}]{Lin:2004nb}%
  \BibitemOpen
  \bibfield  {author} {\bibinfo {author} {\bibfnamefont {H.}~\bibnamefont
  {Lin}}, \bibinfo {author} {\bibfnamefont {O.}~\bibnamefont {Lunin}}, \ and\
  \bibinfo {author} {\bibfnamefont {J.~M.}\ \bibnamefont {Maldacena}},\ }\href
  {\doibase 10.1088/1126-6708/2004/10/025} {\bibfield  {journal} {\bibinfo
  {journal} {JHEP}\ }\textbf {\bibinfo {volume} {0410}},\ \bibinfo {pages}
  {025} (\bibinfo {year} {2004})},\ \Eprint
  {http://arxiv.org/abs/hep-th/0409174} {arXiv:hep-th/0409174}
  \BibitemShut {NoStop}%
%%CITATION = HEP-TH/0409174;%%
\bibitem [{\citenamefont {Gaiotto}\ and\ \citenamefont
  {Maldacena}(2012)}]{Gaiotto:2009gz}%
  \BibitemOpen
  \bibfield  {author} {\bibinfo {author} {\bibfnamefont {D.}~\bibnamefont
  {Gaiotto}}\ and\ \bibinfo {author} {\bibfnamefont {J.}~\bibnamefont
  {Maldacena}},\ }\href {\doibase 10.1007/JHEP10(2012)189} {\bibfield
  {journal} {\bibinfo  {journal} {JHEP}\ }\textbf {\bibinfo {volume} {1210}},\
  \bibinfo {pages} {189} (\bibinfo {year} {2012})},\ \Eprint
  {http://arxiv.org/abs/0904.4466} {arXiv:0904.4466} \BibitemShut
  {NoStop}%
%%CITATION = ARXIV:0904.4466;%%
\bibitem [{\citenamefont {Gadde}\ \emph {et~al.}(2009)\citenamefont {Gadde},
  \citenamefont {Pomoni},\ and\ \citenamefont {Rastelli}}]{Gadde:2009dj}%
  \BibitemOpen
  \bibfield  {author} {\bibinfo {author} {\bibfnamefont {A.}~\bibnamefont
  {Gadde}}, \bibinfo {author} {\bibfnamefont {E.}~\bibnamefont {Pomoni}}, \
  and\ \bibinfo {author} {\bibfnamefont {L.}~\bibnamefont {Rastelli}},\
  }\href@noop {} {\  (\bibinfo {year} {2009})},\ \Eprint
  {http://arxiv.org/abs/0912.4918} {arXiv:0912.4918} \BibitemShut
  {NoStop}%
%%CITATION = ARXIV:0912.4918;%%
\bibitem [{\citenamefont {Reid-Edwards}\ and\ \citenamefont
  {Stefanski}(2011)}]{ReidEdwards:2010qs}%
  \BibitemOpen
  \bibfield  {author} {\bibinfo {author} {\bibfnamefont {R.}~\bibnamefont
  {Reid-Edwards}}\ and\ \bibinfo {author} {\bibfnamefont {B.}~\bibnamefont
  {Stefanski}},\ }\href {\doibase
  10.1016/j.nuclphysb.2011.04.002} {\bibfield  {journal} {\bibinfo  {journal}
  {Nucl.Phys.}\ }\textbf {\bibinfo {volume} {B849}},\ \bibinfo {pages} {549}
  (\bibinfo {year} {2011})},\ \Eprint {http://arxiv.org/abs/1011.0216}
  {arXiv:1011.0216} \BibitemShut {NoStop}%
%%CITATION = ARXIV:1011.0216;%%
\bibitem [{\citenamefont {O~Colgain}\ and\ \citenamefont
  {Stefanski}(2011)}]{Colgain:2011hb}%
  \BibitemOpen
  \bibfield  {author} {\bibinfo {author} {\bibfnamefont {E.}~\bibnamefont
  {O~Colgain}}\ and\ \bibinfo {author} {\bibfnamefont {B.}~\bibnamefont
  {Stefanski}},\ }\href {\doibase
  10.1007/JHEP10(2011)061} {\bibfield  {journal} {\bibinfo  {journal} {JHEP}\
  }\textbf {\bibinfo {volume} {1110}},\ \bibinfo {pages} {061} (\bibinfo {year}
  {2011})},\ \Eprint {http://arxiv.org/abs/1107.5763} {arXiv:1107.5763} \BibitemShut {NoStop}%
%%CITATION = ARXIV:1107.5763;%%
\bibitem [{\citenamefont {Aharony}\ \emph {et~al.}(2012)\citenamefont
  {Aharony}, \citenamefont {Berdichevsky},\ and\ \citenamefont
  {Berkooz}}]{Aharony:2012tz}%
  \BibitemOpen
  \bibfield  {author} {\bibinfo {author} {\bibfnamefont {O.}~\bibnamefont
  {Aharony}}, \bibinfo {author} {\bibfnamefont {L.}~\bibnamefont
  {Berdichevsky}}, \ and\ \bibinfo {author} {\bibfnamefont {M.}~\bibnamefont
  {Berkooz}},\ }\href {\doibase 10.1007/JHEP08(2012)131} {\bibfield  {journal}
  {\bibinfo  {journal} {JHEP}\ }\textbf {\bibinfo {volume} {1208}},\ \bibinfo
  {pages} {131} (\bibinfo {year} {2012})},\ \Eprint
  {http://arxiv.org/abs/1206.5916} {arXiv:1206.5916} \BibitemShut
  {NoStop}%
%%CITATION = ARXIV:1206.5916;%%
\bibitem [{\citenamefont {Stefanski}(2014)}]{Stefanski:2013osa}%
  \BibitemOpen
  \bibfield  {author} {\bibinfo {author} {\bibfnamefont {B.}~\bibnamefont
  {Stefanski}},\ }\href {\doibase 10.1016/j.nuclphysb.2014.03.028} {\bibfield
  {journal} {\bibinfo  {journal} {Nucl.Phys.}\ }\textbf {\bibinfo {volume}
  {B883}},\ \bibinfo {pages} {581} (\bibinfo {year} {2014})},\ \Eprint
  {http://arxiv.org/abs/1308.2789} {arXiv:1308.2789} \BibitemShut
  {NoStop}%
%%CITATION = ARXIV:1308.2789;%%
\bibitem [{\citenamefont {Kachru}\ and\ \citenamefont
  {Silverstein}(1998)}]{Kachru:1998ys}%
  \BibitemOpen
  \bibfield  {author} {\bibinfo {author} {\bibfnamefont {S.}~\bibnamefont
  {Kachru}}\ and\ \bibinfo {author} {\bibfnamefont {E.}~\bibnamefont
  {Silverstein}},\ }\href {\doibase 10.1103/PhysRevLett.80.4855} {\bibfield
  {journal} {\bibinfo  {journal} {Phys.Rev.Lett.}\ }\textbf {\bibinfo {volume}
  {80}},\ \bibinfo {pages} {4855} (\bibinfo {year} {1998})},\ \Eprint
  {http://arxiv.org/abs/hep-th/9802183} {arXiv:hep-th/9802183}
  \BibitemShut {NoStop}%
%%CITATION = HEP-TH/9802183;%%
\bibitem [{\citenamefont {Lawrence}\ \emph {et~al.}(1998)\citenamefont
  {Lawrence}, \citenamefont {Nekrasov},\ and\ \citenamefont
  {Vafa}}]{Lawrence:1998ja}%
  \BibitemOpen
  \bibfield  {author} {\bibinfo {author} {\bibfnamefont {A.~E.}\ \bibnamefont
  {Lawrence}}, \bibinfo {author} {\bibfnamefont {N.}~\bibnamefont {Nekrasov}},
  \ and\ \bibinfo {author} {\bibfnamefont {C.}~\bibnamefont {Vafa}},\ }\href
  {\doibase 10.1016/S0550-3213(98)00495-7} {\bibfield  {journal} {\bibinfo
  {journal} {Nucl.Phys.}\ }\textbf {\bibinfo {volume} {B533}},\ \bibinfo
  {pages} {199} (\bibinfo {year} {1998})},\ \Eprint
  {http://arxiv.org/abs/hep-th/9803015} {arXiv:hep-th/9803015}
  \BibitemShut {NoStop}%
%%CITATION = HEP-TH/9803015;%%
\bibitem [{\citenamefont {Gukov}(1998)}]{Gukov:1998kk}%
  \BibitemOpen
  \bibfield  {author} {\bibinfo {author} {\bibfnamefont {S.}~\bibnamefont
  {Gukov}},\ }\href {\doibase 10.1016/S0370-2693(98)01005-3} {\bibfield
  {journal} {\bibinfo  {journal} {Phys.Lett.}\ }\textbf {\bibinfo {volume}
  {B439}},\ \bibinfo {pages} {23} (\bibinfo {year} {1998})},\ \Eprint
  {http://arxiv.org/abs/hep-th/9806180} {arXiv:hep-th/9806180}
  \BibitemShut {NoStop}%
%%CITATION = HEP-TH/9806180;%%
\bibitem [{\citenamefont {Erickson}\ \emph {et~al.}(2000)\citenamefont
  {Erickson}, \citenamefont {Semenoff},\ and\ \citenamefont
  {Zarembo}}]{Erickson:2000af}%
  \BibitemOpen
  \bibfield  {author} {\bibinfo {author} {\bibfnamefont {J.}~\bibnamefont
  {Erickson}}, \bibinfo {author} {\bibfnamefont {G.}~\bibnamefont {Semenoff}},
  \ and\ \bibinfo {author} {\bibfnamefont {K.}~\bibnamefont {Zarembo}},\ }\href
  {\doibase 10.1016/S0550-3213(00)00300-X} {\bibfield  {journal} {\bibinfo
  {journal} {Nucl.Phys.}\ }\textbf {\bibinfo {volume} {B582}},\ \bibinfo
  {pages} {155} (\bibinfo {year} {2000})},\ \Eprint
  {http://arxiv.org/abs/hep-th/0003055} {arXiv:hep-th/0003055}
  \BibitemShut {NoStop}%
%%CITATION = HEP-TH/0003055;%%
\bibitem [{\citenamefont {Drukker}\ and\ \citenamefont
  {Gross}(2001)}]{Drukker:2000rr}%
  \BibitemOpen
  \bibfield  {author} {\bibinfo {author} {\bibfnamefont {N.}~\bibnamefont
  {Drukker}}\ and\ \bibinfo {author} {\bibfnamefont {D.~J.}\ \bibnamefont
  {Gross}},\ }\href {\doibase 10.1063/1.1372177} {\bibfield  {journal}
  {\bibinfo  {journal} {J.Math.Phys.}\ }\textbf {\bibinfo {volume} {42}},\
  \bibinfo {pages} {2896} (\bibinfo {year} {2001})},\ \Eprint
  {http://arxiv.org/abs/hep-th/0010274} {arXiv:hep-th/0010274}
  \BibitemShut {NoStop}%
%%CITATION = HEP-TH/0010274;%%
\bibitem [{\citenamefont {Howe}\ \emph {et~al.}(1983)\citenamefont {Howe},
  \citenamefont {Stelle},\ and\ \citenamefont {West}}]{Howe:1983wj}%
  \BibitemOpen
  \bibfield  {author} {\bibinfo {author} {\bibfnamefont {P.~S.}\ \bibnamefont
  {Howe}}, \bibinfo {author} {\bibfnamefont {K.}~\bibnamefont {Stelle}}, \ and\
  \bibinfo {author} {\bibfnamefont {P.~C.}\ \bibnamefont {West}},\ }\href
  {\doibase 10.1016/0370-2693(83)91402-8} {\bibfield  {journal} {\bibinfo
  {journal} {Phys.Lett.}\ }\textbf {\bibinfo {volume} {B124}},\ \bibinfo
  {pages} {55} (\bibinfo {year} {1983})}\BibitemShut {NoStop}%
%%CITATION = PHLTA,B124,55;%%
\bibitem [{\citenamefont {Katz}\ \emph {et~al.}(1998)\citenamefont {Katz},
  \citenamefont {Mayr},\ and\ \citenamefont {Vafa}}]{Katz:1997eq}%
  \BibitemOpen
  \bibfield  {author} {\bibinfo {author} {\bibfnamefont {S.}~\bibnamefont
  {Katz}}, \bibinfo {author} {\bibfnamefont {P.}~\bibnamefont {Mayr}}, \ and\
  \bibinfo {author} {\bibfnamefont {C.}~\bibnamefont {Vafa}},\ }\href@noop {}
  {\bibfield  {journal} {\bibinfo  {journal} {Adv.Theor.Math.Phys.}\ }\textbf
  {\bibinfo {volume} {1}},\ \bibinfo {pages} {53} (\bibinfo {year} {1998})},\
  \Eprint {http://arxiv.org/abs/hep-th/9706110} {arXiv:hep-th/9706110}
  \BibitemShut {NoStop}%
%%CITATION = HEP-TH/9706110;%%
\bibitem [{\citenamefont {Nekrasov}\ and\ \citenamefont
  {Pestun}(2012)}]{Nekrasov:2012xe}%
  \BibitemOpen
  \bibfield  {author} {\bibinfo {author} {\bibfnamefont {N.}~\bibnamefont
  {Nekrasov}}\ and\ \bibinfo {author} {\bibfnamefont {V.}~\bibnamefont
  {Pestun}},\ }\href@noop {} {\  (\bibinfo {year} {2012})},\ \Eprint
  {http://arxiv.org/abs/1211.2240} {arXiv:1211.2240} \BibitemShut
  {NoStop}%
%%CITATION = ARXIV:1211.2240;%%
\bibitem [{\citenamefont {Gadde}\ \emph {et~al.}(2012)\citenamefont {Gadde},
  \citenamefont {Pomoni},\ and\ \citenamefont {Rastelli}}]{Gadde:2010zi}%
  \BibitemOpen
  \bibfield  {author} {\bibinfo {author} {\bibfnamefont {A.}~\bibnamefont
  {Gadde}}, \bibinfo {author} {\bibfnamefont {E.}~\bibnamefont {Pomoni}}, \
  and\ \bibinfo {author} {\bibfnamefont {L.}~\bibnamefont {Rastelli}},\ }\href
  {\doibase 10.1007/JHEP06(2012)107} {\bibfield  {journal} {\bibinfo  {journal}
  {JHEP}\ }\textbf {\bibinfo {volume} {1206}},\ \bibinfo {pages} {107}
  (\bibinfo {year} {2012})},\ \Eprint {http://arxiv.org/abs/1006.0015}
  {arXiv:1006.0015} \BibitemShut {NoStop}%
%%CITATION = ARXIV:1006.0015;%%
\bibitem [{\citenamefont {Gadde}\ and\ \citenamefont
  {Rastelli}(2012)}]{Gadde:2010ku}%
  \BibitemOpen
  \bibfield  {author} {\bibinfo {author} {\bibfnamefont {A.}~\bibnamefont
  {Gadde}}\ and\ \bibinfo {author} {\bibfnamefont {L.}~\bibnamefont
  {Rastelli}},\ }\href {\doibase 10.1007/JHEP04(2012)053} {\bibfield  {journal}
  {\bibinfo  {journal} {JHEP}\ }\textbf {\bibinfo {volume} {1204}},\ \bibinfo
  {pages} {053} (\bibinfo {year} {2012})},\ \Eprint
  {http://arxiv.org/abs/1012.2097} {arXiv:1012.2097} \BibitemShut
  {NoStop}%
%%CITATION = ARXIV:1012.2097;%%
\bibitem [{\citenamefont {Pomoni}\ and\ \citenamefont
  {Sieg}(2011)}]{Pomoni:2011jj}%
  \BibitemOpen
  \bibfield  {author} {\bibinfo {author} {\bibfnamefont {E.}~\bibnamefont
  {Pomoni}}\ and\ \bibinfo {author} {\bibfnamefont {C.}~\bibnamefont {Sieg}},\
  }\href@noop {} {\  (\bibinfo {year} {2011})},\ \Eprint
  {http://arxiv.org/abs/1105.3487} {arXiv:1105.3487} \BibitemShut
  {NoStop}%
%%CITATION = ARXIV:1105.3487;%%
\bibitem [{\citenamefont {Liendo}\ \emph {et~al.}(2012)\citenamefont {Liendo},
  \citenamefont {Pomoni},\ and\ \citenamefont {Rastelli}}]{Liendo:2011xb}%
  \BibitemOpen
  \bibfield  {author} {\bibinfo {author} {\bibfnamefont {P.}~\bibnamefont
  {Liendo}}, \bibinfo {author} {\bibfnamefont {E.}~\bibnamefont {Pomoni}}, \
  and\ \bibinfo {author} {\bibfnamefont {L.}~\bibnamefont {Rastelli}},\ }\href
  {\doibase 10.1007/JHEP07(2012)003} {\bibfield  {journal} {\bibinfo  {journal}
  {JHEP}\ }\textbf {\bibinfo {volume} {1207}},\ \bibinfo {pages} {003}
  (\bibinfo {year} {2012})},\ \Eprint {http://arxiv.org/abs/1105.3972}
  {arXiv:1105.3972} \BibitemShut {NoStop}%
%%CITATION = ARXIV:1105.3972;%%
\bibitem [{\citenamefont {Gadde}\ \emph {et~al.}(2013)\citenamefont {Gadde},
  \citenamefont {Liendo}, \citenamefont {Rastelli},\ and\ \citenamefont
  {Yan}}]{Gadde:2012rv}%
  \BibitemOpen
  \bibfield  {author} {\bibinfo {author} {\bibfnamefont {A.}~\bibnamefont
  {Gadde}}, \bibinfo {author} {\bibfnamefont {P.}~\bibnamefont {Liendo}},
  \bibinfo {author} {\bibfnamefont {L.}~\bibnamefont {Rastelli}}, \ and\
  \bibinfo {author} {\bibfnamefont {W.}~\bibnamefont {Yan}},\ }\href {\doibase
  10.1007/JHEP08(2013)015} {\bibfield  {journal} {\bibinfo  {journal} {JHEP}\
  }\textbf {\bibinfo {volume} {1308}},\ \bibinfo {pages} {015} (\bibinfo {year}
  {2013})},\ \Eprint {http://arxiv.org/abs/1211.0271} {arXiv:1211.0271}
  \BibitemShut {NoStop}%
%%CITATION = ARXIV:1211.0271;%%
\bibitem [{\citenamefont {Andree}\ and\ \citenamefont
  {Young}(2010)}]{Andree:2010na}%
  \BibitemOpen
  \bibfield  {author} {\bibinfo {author} {\bibfnamefont {R.}~\bibnamefont
  {Andree}}\ and\ \bibinfo {author} {\bibfnamefont {D.}~\bibnamefont {Young}},\
  }\href {\doibase 10.1007/JHEP09(2010)095} {\bibfield  {journal} {\bibinfo
  {journal} {JHEP}\ }\textbf {\bibinfo {volume} {1009}},\ \bibinfo {pages}
  {095} (\bibinfo {year} {2010})},\ \Eprint {http://arxiv.org/abs/1007.4923}
  {arXiv:1007.4923} \BibitemShut {NoStop}%
%%CITATION = ARXIV:1007.4923;%%
\bibitem [{\citenamefont {Gates}\ \emph {et~al.}(1983)\citenamefont {Gates},
  \citenamefont {Grisaru}, \citenamefont {Rocek},\ and\ \citenamefont
  {Siegel}}]{Gates:1983nr}%
  \BibitemOpen
  \bibfield  {author} {\bibinfo {author} {\bibfnamefont {S.}~\bibnamefont
  {Gates}}, \bibinfo {author} {\bibfnamefont {M.~T.}\ \bibnamefont {Grisaru}},
  \bibinfo {author} {\bibfnamefont {M.}~\bibnamefont {Rocek}}, \ and\ \bibinfo
  {author} {\bibfnamefont {W.}~\bibnamefont {Siegel}},\ }\href@noop {} {\
  (\bibinfo {year} {1983})},\ \Eprint {http://arxiv.org/abs/hep-th/0108200}
  {arXiv:hep-th/0108200} \BibitemShut {NoStop}%
%%CITATION = HEP-TH/0108200;%%
\bibitem [{\citenamefont {Buchbinder}\ \emph {et~al.}(1998)\citenamefont
  {Buchbinder}, \citenamefont {Buchbinder}, \citenamefont {Kuzenko},\ and\
  \citenamefont {Ovrut}}]{Buchbinder:1997ya}%
  \BibitemOpen
  \bibfield  {author} {\bibinfo {author} {\bibfnamefont {I.}~\bibnamefont
  {Buchbinder}}, \bibinfo {author} {\bibfnamefont {E.}~\bibnamefont
  {Buchbinder}}, \bibinfo {author} {\bibfnamefont {S.}~\bibnamefont {Kuzenko}},
  \ and\ \bibinfo {author} {\bibfnamefont {B.~A.}\ \bibnamefont {Ovrut}},\
  }\href {\doibase 10.1016/S0370-2693(97)01319-1} {\bibfield  {journal}
  {\bibinfo  {journal} {Phys.Lett.}\ }\textbf {\bibinfo {volume} {B417}},\
  \bibinfo {pages} {61} (\bibinfo {year} {1998})},\ \Eprint
  {http://arxiv.org/abs/hep-th/9704214} {arXiv:hep-th/9704214}
  \BibitemShut {NoStop}%
%%CITATION = HEP-TH/9704214;%%
\bibitem [{\citenamefont {Buchbinder}\ and\ \citenamefont
  {Kuzenko}(1998)}]{Buchbinder:1998np}%
  \BibitemOpen
  \bibfield  {author} {\bibinfo {author} {\bibfnamefont {I.~L.}\ \bibnamefont
  {Buchbinder}}\ and\ \bibinfo {author} {\bibfnamefont {S.~M.}\ \bibnamefont
  {Kuzenko}},\ }\href {\doibase 10.1142/S0217732398001704} {\bibfield
  {journal} {\bibinfo  {journal} {Mod.Phys.Lett.}\ }\textbf {\bibinfo {volume}
  {A13}},\ \bibinfo {pages} {1623} (\bibinfo {year} {1998})},\ \Eprint
  {http://arxiv.org/abs/hep-th/9804168} {arXiv:hep-th/9804168}
  \BibitemShut {NoStop}%
%%CITATION = HEP-TH/9804168;%%
\bibitem [{\citenamefont {Jain}\ and\ \citenamefont
  {Siegel}(2013)}]{Jain:2013hua}%
  \BibitemOpen
  \bibfield  {author} {\bibinfo {author} {\bibfnamefont {D.}~\bibnamefont
  {Jain}}\ and\ \bibinfo {author} {\bibfnamefont {W.}~\bibnamefont {Siegel}},\
  }\href {\doibase 10.1103/PhysRevD.88.025018} {\bibfield  {journal} {\bibinfo
  {journal} {Phys.Rev.}\ }\textbf {\bibinfo {volume} {D88}},\ \bibinfo {pages}
  {025018} (\bibinfo {year} {2013})},\ \Eprint {http://arxiv.org/abs/1302.3277}
  {arXiv:1302.3277} \BibitemShut {NoStop}%
%%CITATION = ARXIV:1302.3277;%%
\bibitem [{\citenamefont {Fiamberti}\ \emph
  {et~al.}(2008{\natexlab{a}})\citenamefont {Fiamberti}, \citenamefont
  {Santambrogio}, \citenamefont {Sieg},\ and\ \citenamefont
  {Zanon}}]{Fiamberti:2008sh}%
  \BibitemOpen
  \bibfield  {author} {\bibinfo {author} {\bibfnamefont {F.}~\bibnamefont
  {Fiamberti}}, \bibinfo {author} {\bibfnamefont {A.}~\bibnamefont
  {Santambrogio}}, \bibinfo {author} {\bibfnamefont {C.}~\bibnamefont {Sieg}},
  \ and\ \bibinfo {author} {\bibfnamefont {D.}~\bibnamefont {Zanon}},\ }\href
  {\doibase 10.1016/j.nuclphysb.2008.07.014} {\bibfield  {journal} {\bibinfo
  {journal} {Nucl.Phys.}\ }\textbf {\bibinfo {volume} {B805}},\ \bibinfo
  {pages} {231} (\bibinfo {year} {2008}{\natexlab{a}})},\ \Eprint
  {http://arxiv.org/abs/0806.2095} {arXiv:0806.2095} \BibitemShut
  {NoStop}%
%%CITATION = ARXIV:0806.2095;%%
\bibitem [{\citenamefont {Sieg}(2011)}]{Sieg:2010tz}%
  \BibitemOpen
  \bibfield  {author} {\bibinfo {author} {\bibfnamefont {C.}~\bibnamefont
  {Sieg}},\ }\href {\doibase 10.1103/PhysRevD.84.045014} {\bibfield  {journal}
  {\bibinfo  {journal} {Phys.Rev.}\ }\textbf {\bibinfo {volume} {D84}},\
  \bibinfo {pages} {045014} (\bibinfo {year} {2011})},\ \Eprint
  {http://arxiv.org/abs/1008.3351} {arXiv:1008.3351} \BibitemShut
  {NoStop}%
%%CITATION = ARXIV:1008.3351;%%
\bibitem [{\citenamefont {Bershadsky}\ \emph {et~al.}(1998)\citenamefont
  {Bershadsky}, \citenamefont {Kakushadze},\ and\ \citenamefont
  {Vafa}}]{Bershadsky:1998mb}%
  \BibitemOpen
  \bibfield  {author} {\bibinfo {author} {\bibfnamefont {M.}~\bibnamefont
  {Bershadsky}}, \bibinfo {author} {\bibfnamefont {Z.}~\bibnamefont
  {Kakushadze}}, \ and\ \bibinfo {author} {\bibfnamefont {C.}~\bibnamefont
  {Vafa}},\ }\href {\doibase 10.1016/S0550-3213(98)00272-7} {\bibfield
  {journal} {\bibinfo  {journal} {Nucl.Phys.}\ }\textbf {\bibinfo {volume}
  {B523}},\ \bibinfo {pages} {59} (\bibinfo {year} {1998})},\ \Eprint
  {http://arxiv.org/abs/hep-th/9803076} {arXiv:hep-th/9803076}
  \BibitemShut {NoStop}%
%%CITATION = HEP-TH/9803076;%%
\bibitem [{\citenamefont {Bershadsky}\ and\ \citenamefont
  {Johansen}(1998)}]{Bershadsky:1998cb}%
  \BibitemOpen
  \bibfield  {author} {\bibinfo {author} {\bibfnamefont {M.}~\bibnamefont
  {Bershadsky}}\ and\ \bibinfo {author} {\bibfnamefont {A.}~\bibnamefont
  {Johansen}},\ }\href {\doibase 10.1016/S0550-3213(98)00526-4} {\bibfield
  {journal} {\bibinfo  {journal} {Nucl.Phys.}\ }\textbf {\bibinfo {volume}
  {B536}},\ \bibinfo {pages} {141} (\bibinfo {year} {1998})},\ \Eprint
  {http://arxiv.org/abs/hep-th/9803249} {arXiv:hep-th/9803249}
  \BibitemShut {NoStop}%
%%CITATION = HEP-TH/9803249;%%
\bibitem [{\citenamefont {Passerini}\ and\ \citenamefont
  {Zarembo}(2011)}]{Passerini:2011fe}%
  \BibitemOpen
  \bibfield  {author} {\bibinfo {author} {\bibfnamefont {F.}~\bibnamefont
  {Passerini}}\ and\ \bibinfo {author} {\bibfnamefont {K.}~\bibnamefont
  {Zarembo}},\ }\href {\doibase 10.1007/JHEP10(2011)065,
  10.1007/JHEP09(2011)102} {\bibfield  {journal} {\bibinfo  {journal} {JHEP}\
  }\textbf {\bibinfo {volume} {1109}},\ \bibinfo {pages} {102} (\bibinfo {year}
  {2011})},\ \Eprint {http://arxiv.org/abs/1106.5763} {arXiv:1106.5763} \BibitemShut {NoStop}%
%%CITATION = ARXIV:1106.5763;%%
\bibitem [{\citenamefont {Russo}\ and\ \citenamefont
  {Zarembo}(2012)}]{Russo:2012ay}%
  \BibitemOpen
  \bibfield  {author} {\bibinfo {author} {\bibfnamefont {J.}~\bibnamefont
  {Russo}}\ and\ \bibinfo {author} {\bibfnamefont {K.}~\bibnamefont
  {Zarembo}},\ }\href {\doibase 10.1007/JHEP10(2012)082} {\bibfield  {journal}
  {\bibinfo  {journal} {JHEP}\ }\textbf {\bibinfo {volume} {1210}},\ \bibinfo
  {pages} {082} (\bibinfo {year} {2012})},\ \Eprint
  {http://arxiv.org/abs/1207.3806} {arXiv:1207.3806} \BibitemShut
  {NoStop}%
%%CITATION = ARXIV:1207.3806;%%
\bibitem [{\citenamefont {Russo}\ and\ \citenamefont
  {Zarembo}(2013{\natexlab{a}})}]{Russo:2013kea}%
  \BibitemOpen
  \bibfield  {author} {\bibinfo {author} {\bibfnamefont {J.}~\bibnamefont
  {Russo}}\ and\ \bibinfo {author} {\bibfnamefont {K.}~\bibnamefont
  {Zarembo}},\ }\href {\doibase 10.1007/JHEP11(2013)130} {\bibfield  {journal}
  {\bibinfo  {journal} {JHEP}\ }\textbf {\bibinfo {volume} {1311}},\ \bibinfo
  {pages} {130} (\bibinfo {year} {2013}{\natexlab{a}})},\ \Eprint
  {http://arxiv.org/abs/1309.1004} {arXiv:1309.1004} \BibitemShut
  {NoStop}%
%%CITATION = ARXIV:1309.1004;%%
\bibitem [{\citenamefont {Russo}\ and\ \citenamefont
  {Zarembo}(2013{\natexlab{b}})}]{Russo:2013sba}%
  \BibitemOpen
  \bibfield  {author} {\bibinfo {author} {\bibfnamefont {J.}~\bibnamefont
  {Russo}}\ and\ \bibinfo {author} {\bibfnamefont {K.}~\bibnamefont
  {Zarembo}},\ }\href@noop {} {\  (\bibinfo {year} {2013}{\natexlab{b}})},\
  \Eprint {http://arxiv.org/abs/1312.1214} {arXiv:1312.1214}
  \BibitemShut {NoStop}%
%%CITATION = ARXIV:1312.1214;%%
\bibitem [{\citenamefont {Broadhurst}(1985)}]{Broadhurst:1985vq}%
  \BibitemOpen
  \bibfield  {author} {\bibinfo {author} {\bibfnamefont {D.~J.}\ \bibnamefont
  {Broadhurst}},\ }\href {\doibase 10.1016/0370-2693(85)90340-5} {\bibfield
  {journal} {\bibinfo  {journal} {Phys.Lett.}\ }\textbf {\bibinfo {volume}
  {B164}},\ \bibinfo {pages} {356} (\bibinfo {year} {1985})}\BibitemShut
  {NoStop}%
%%CITATION = PHLTA,B164,356;%%
\bibitem [{\citenamefont {Korchemsky}(2012)}]{Korchemsky:2010kj}%
  \BibitemOpen
  \bibfield  {author} {\bibinfo {author} {\bibfnamefont {G.}~\bibnamefont
  {Korchemsky}},\ }\href {\doibase 10.1007/s11005-011-0516-7} {\bibfield
  {journal} {\bibinfo  {journal} {Lett.Math.Phys.}\ }\textbf {\bibinfo {volume}
  {99}},\ \bibinfo {pages} {425} (\bibinfo {year} {2012})},\ \Eprint
  {http://arxiv.org/abs/1012.4000} {arXiv:1012.4000} \BibitemShut
  {NoStop}%
%%CITATION = ARXIV:1012.4000;%%
\bibitem [{\citenamefont {Caron-Huot}(2011)}]{CaronHuot:2010ek}%
  \BibitemOpen
  \bibfield  {author} {\bibinfo {author} {\bibfnamefont {S.}~\bibnamefont
  {Caron-Huot}},\ }\href {\doibase 10.1007/JHEP07(2011)058} {\bibfield
  {journal} {\bibinfo  {journal} {JHEP}\ }\textbf {\bibinfo {volume} {1107}},\
  \bibinfo {pages} {058} (\bibinfo {year} {2011})},\ \Eprint
  {http://arxiv.org/abs/1010.1167} {arXiv:1010.1167} \BibitemShut
  {NoStop}%
%%CITATION = ARXIV:1010.1167;%%
\bibitem [{\citenamefont {Eden}\ \emph
  {et~al.}(2013{\natexlab{a}})\citenamefont {Eden}, \citenamefont {Heslop},
  \citenamefont {Korchemsky},\ and\ \citenamefont {Sokatchev}}]{Eden:2011yp}%
  \BibitemOpen
  \bibfield  {author} {\bibinfo {author} {\bibfnamefont {B.}~\bibnamefont
  {Eden}}, \bibinfo {author} {\bibfnamefont {P.}~\bibnamefont {Heslop}},
  \bibinfo {author} {\bibfnamefont {G.~P.}\ \bibnamefont {Korchemsky}}, \ and\
  \bibinfo {author} {\bibfnamefont {E.}~\bibnamefont {Sokatchev}},\ }\href
  {\doibase 10.1016/j.nuclphysb.2012.12.015} {\bibfield  {journal} {\bibinfo
  {journal} {Nucl.Phys.}\ }\textbf {\bibinfo {volume} {B869}},\ \bibinfo
  {pages} {329} (\bibinfo {year} {2013}{\natexlab{a}})},\ \Eprint
  {http://arxiv.org/abs/1103.3714} {arXiv:1103.3714} \BibitemShut
  {NoStop}%
%%CITATION = ARXIV:1103.3714;%%
\bibitem [{\citenamefont {Eden}\ \emph
  {et~al.}(2013{\natexlab{b}})\citenamefont {Eden}, \citenamefont {Heslop},
  \citenamefont {Korchemsky},\ and\ \citenamefont {Sokatchev}}]{Eden:2011ku}%
  \BibitemOpen
  \bibfield  {author} {\bibinfo {author} {\bibfnamefont {B.}~\bibnamefont
  {Eden}}, \bibinfo {author} {\bibfnamefont {P.}~\bibnamefont {Heslop}},
  \bibinfo {author} {\bibfnamefont {G.~P.}\ \bibnamefont {Korchemsky}}, \ and\
  \bibinfo {author} {\bibfnamefont {E.}~\bibnamefont {Sokatchev}},\ }\href
  {\doibase 10.1016/j.nuclphysb.2012.12.014} {\bibfield  {journal} {\bibinfo
  {journal} {Nucl.Phys.}\ }\textbf {\bibinfo {volume} {B869}},\ \bibinfo
  {pages} {378} (\bibinfo {year} {2013}{\natexlab{b}})},\ \Eprint
  {http://arxiv.org/abs/1103.4353} {arXiv:1103.4353} \BibitemShut
  {NoStop}%
%%CITATION = ARXIV:1103.4353;%%
\bibitem [{\citenamefont {Leurent}\ and\ \citenamefont
  {Volin}(2013)}]{Leurent:2013mr}%
  \BibitemOpen
  \bibfield  {author} {\bibinfo {author} {\bibfnamefont {S.}~\bibnamefont
  {Leurent}}\ and\ \bibinfo {author} {\bibfnamefont {D.}~\bibnamefont
  {Volin}},\ }\href {\doibase 10.1016/j.nuclphysb.2013.07.020} {\bibfield
  {journal} {\bibinfo  {journal} {Nucl.Phys.}\ }\textbf {\bibinfo {volume}
  {B875}},\ \bibinfo {pages} {757} (\bibinfo {year} {2013})},\ \Eprint
  {http://arxiv.org/abs/1302.1135} {arXiv:1302.1135} \BibitemShut
  {NoStop}%
%%CITATION = ARXIV:1302.1135;%%
\bibitem [{\citenamefont {Fiamberti}\ \emph
  {et~al.}(2008{\natexlab{b}})\citenamefont {Fiamberti}, \citenamefont
  {Santambrogio}, \citenamefont {Sieg},\ and\ \citenamefont
  {Zanon}}]{Fiamberti:2007rj}%
  \BibitemOpen
  \bibfield  {author} {\bibinfo {author} {\bibfnamefont {F.}~\bibnamefont
  {Fiamberti}}, \bibinfo {author} {\bibfnamefont {A.}~\bibnamefont
  {Santambrogio}}, \bibinfo {author} {\bibfnamefont {C.}~\bibnamefont {Sieg}},
  \ and\ \bibinfo {author} {\bibfnamefont {D.}~\bibnamefont {Zanon}},\ }\href
  {\doibase 10.1016/j.physletb.2008.06.061} {\bibfield  {journal} {\bibinfo
  {journal} {Phys.Lett.}\ }\textbf {\bibinfo {volume} {B666}},\ \bibinfo
  {pages} {100} (\bibinfo {year} {2008}{\natexlab{b}})},\ \Eprint
  {http://arxiv.org/abs/0712.3522} {arXiv:0712.3522} \BibitemShut
  {NoStop}%
%%CITATION = ARXIV:0712.3522;%%
\bibitem [{\citenamefont {Gromov}\ and\ \citenamefont
  {Sizov}(2014)}]{Gromov:2014eha}%
  \BibitemOpen
  \bibfield  {author} {\bibinfo {author} {\bibfnamefont {N.}~\bibnamefont
  {Gromov}}\ and\ \bibinfo {author} {\bibfnamefont {G.}~\bibnamefont {Sizov}},\
  }\href@noop {} {\  (\bibinfo {year} {2014})},\ \Eprint
  {http://arxiv.org/abs/1403.1894} {arXiv:1403.1894} \BibitemShut
  {NoStop}%
%%CITATION = ARXIV:1403.1894;%%
\bibitem [{\citenamefont {Basso}(2011)}]{Basso:2011rs}%
  \BibitemOpen
  \bibfield  {author} {\bibinfo {author} {\bibfnamefont {B.}~\bibnamefont
  {Basso}},\ }\href@noop {} {\  (\bibinfo {year} {2011})},\ \Eprint
  {http://arxiv.org/abs/1109.3154} {arXiv:1109.3154} \BibitemShut
  {NoStop}%
%%CITATION = ARXIV:1109.3154;%%
\bibitem [{\citenamefont {Aharony}\ \emph {et~al.}(2008)\citenamefont
  {Aharony}, \citenamefont {Bergman},\ and\ \citenamefont
  {Jafferis}}]{Aharony:2008gk}%
  \BibitemOpen
  \bibfield  {author} {\bibinfo {author} {\bibfnamefont {O.}~\bibnamefont
  {Aharony}}, \bibinfo {author} {\bibfnamefont {O.}~\bibnamefont {Bergman}}, \
  and\ \bibinfo {author} {\bibfnamefont {D.~L.}\ \bibnamefont {Jafferis}},\
  }\href {\doibase 10.1088/1126-6708/2008/11/043} {\bibfield  {journal}
  {\bibinfo  {journal} {JHEP}\ }\textbf {\bibinfo {volume} {0811}},\ \bibinfo
  {pages} {043} (\bibinfo {year} {2008})},\ \Eprint
  {http://arxiv.org/abs/0807.4924} {arXiv:0807.4924} \BibitemShut
  {NoStop}%
%%CITATION = ARXIV:0807.4924;%%
\end{thebibliography}
\end{document}